\documentclass[]{interact}

\usepackage[caption=false]{subfig}

\usepackage[numbers,sort&compress]{natbib}

\usepackage{amsmath,amscd}
\usepackage{amsfonts}
\usepackage{amssymb}
\usepackage{graphicx}
\usepackage{fancyvrb}
\usepackage{xcolor}
\usepackage{stackengine}
\usepackage{bbm}

\usepackage[draft,firstpage]{draftmark}
\draftmarksetup{fontfamily=ptm,angle=0,scale=0.07,mark={\parbox{260cm}{
This is an Accepted Manuscript of an article published by Taylor \& Francis Group in the \emph{International Journal
of Parallel, Emergent \& Distributed Systems} on 16/10/2024, available online: https://doi.org/10.1080/17445760.2024.2413509
}},color=red,xcoord=-78,ycoord=120}

\bibpunct[, ]{[}{]}{,}{n}{,}{,}
\makeatletter
\def\NAT@def@citea{\def\@citea{\NAT@separator}}
\makeatother

\theoremstyle{plain}
\newtheorem{theorem}{Theorem}[section]

\newtheorem{proposition}[theorem]{Proposition}

\theoremstyle{definition}

\theoremstyle{remark}

\newcommand*\chboard{
\begin{smallmatrix}
\square \mkern-9mu & \blacksquare \\[-0.2em]
\blacksquare \mkern-9mu & \square
\end{smallmatrix}
}

\newenvironment{EqVerbatim}
 {\VerbatimEnvironment\begin{gathered}\normalbaselines\begin{BVerbatim}}
 {\end{BVerbatim}\end{gathered}}

\begin{document}


\title{Cellular automaton model of self-healing}

\author{
\name{Henryk Fuk\'s\textsuperscript{a}\thanks{CONTACT Henryk Fuk\'s. Email: hfuks@brocku.ca} and Jos\'e Manuel G\'omez Soto\textsuperscript{b}}
\affil{\textsuperscript{a}Department of Mathematics,
Brock University, St. Catharines, ON, Canada; \textsuperscript{b} Unidad Acad\'emica de Matem\'aticas, 
Universidad Aut\'onoma de Zacatecas,
Calzada Solidaridad entronque Paseo a la Bufa,
Zacatecas, Zac. M\'exico}
}

\maketitle

\begin{abstract}
We propose a simple cellular automaton model of a self-healing 
system and investigate its properties. In the model, the substrate is a two-dimensional checkerboard configuration
which can be damaged by changing values of a finite number of sites.
The cellular automaton we consider is a checkerboard voting rule,
a binary rule with Moore neighbourhood which is topologically
conjugate to majority voting rule. For a single color damage (when only cells in the
same state are modified), the rule always fixes the damage. 
For a general damage, when it is localized inside
a  $3 \times 3$ square, the rule also fixes it always. When the damage is inside of a larger $n \times n$ square, the efficiency of the rule in fixing the damage becomes smaller than $100\%$,  but it remains better than $98\%$ for $n \leq 5$
and better than $75 \%$ for $n\leq 7$. We show that in the limit of infinite $n$ the efficiency  tends to zero. 
\end{abstract}


\section*{Introduction}


Self-healing materials are  synthetically created substances that have the ability to automatically repair damages to themselves.  Polymers, elastomers and other materials exist which have that kind of ability \cite{KELLER2018431}. They are used in many
commercial products, for example, self-healing corrosion protection coatings, self-healing cutting mats or self-healing targets for shooting practice.
A somewhat similar process,  called ``fault-local distributed mending'' \cite{Kutten99a,Kutten99b,Peleg2002}, has been studied
in the context of distributed networks.

In this paper we will explore cellular automata (CA) models of self-healing. 
These are CA which preserve certain periodic patterns, and when the pattern is 
locally damaged, their time evolution leads to the ``repair'' of the damage.
If the pattern is homogeneous, such CA trivially exist. For example,
zero rule (mapping all neighbourhoods to 0) will instantly repair any
damage to the pattern consisting of all zeros.

\begin{figure}
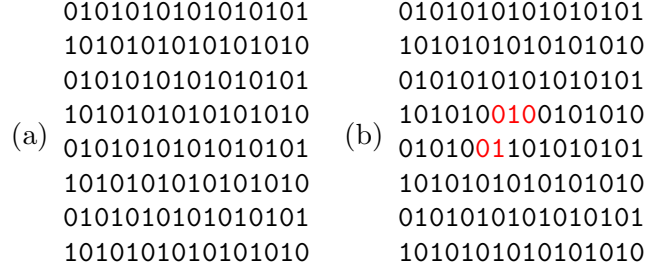

\begin{center}
(a)\,\,\,$
\begin{EqVerbatim}[commandchars=\\\{\}]
0101010101010101
1010101010101010
0101010101010101
1010101010101010
0101010101010101
1010101010101010
0101010101010101
1010101010101010
\end{EqVerbatim}
$ \,\,\,
(b)\,\,\,$
\begin{EqVerbatim}[commandchars=\\\{\}]
0101010101010101
1010101010101010
0101010101010101
101010\textcolor{red}{0}\textcolor{red}{1}\textcolor{red}{0}0101010
01010\textcolor{red}{0}\textcolor{red}{1}101010101
1010101010101010
0101010101010101
1010101010101010
\end{EqVerbatim}
$
\end{center}\caption{Fragment of a perfect checkerboard patern  (a) and  damaged checkerboard pattern (b).} \label{damageexample}
\end{figure}

In two dimensions, the simple non-trivial periodic pattern could be the 
checkerboard pattern, like the one shown in Figure \ref{damageexample}a.
Let us do some ``damage'' to this pattern, for example, like in Figure 
\ref{damageexample}b.
What we want is a binary CA rule which after a finite number of steps, starting from 
pattern of Figure 
\ref{damageexample}b, eventually produces pattern of Figure 
\ref{damageexample}a.

After some experimenting, we found a simple rule  which works pretty well
in healing checkerboard patterns, we will
 call it ``checkerboard voting rule''. It uses Moore neighbourhood.

Let $H(a,b)$ be a Hamming distance, meaning the number of bits on which
$a$ and $b$ are different. The checkerboard voting rule is defined as 
follows:

$$
  f\begin{pmatrix}
     nw & n & ne\\
    w & c & e\\
    sw & s & se
  \end{pmatrix}  
  =\begin{cases}
  1 & 
  \text{if\,\,\,}
  H \left(\begin{smallmatrix}
       nw & n & ne\\
    w & c & e\\
    sw & s & se
  \end{smallmatrix},
  \begin{smallmatrix}
    1 & 0 & 1\\
    0 & 1 & 0\\
    1 & 0 & 1
  \end{smallmatrix}  \right) <
  H \left(\begin{smallmatrix}
       nw & n & ne\\
    w & c & e\\
    sw & s & se
  \end{smallmatrix},
  \begin{smallmatrix}
    0 & 1 & 0\\
    1 & 0 & 1\\
    0 & 1 & 0
  \end{smallmatrix}  \right),
   \\[1em]
  0 & 
  \text{if\,\,\,}
  H \left(\begin{smallmatrix}
       nw & n & ne\\
    w & c & e\\
    sw & s & se
  \end{smallmatrix},
  \begin{smallmatrix}
    1 & 0 & 1\\
    0 & 1 & 0\\
    1 & 0 & 1
  \end{smallmatrix}  \right) >
  H \left(\begin{smallmatrix}
       nw & n & ne\\
    w & c & e\\
    sw & s & se
  \end{smallmatrix},
  \begin{smallmatrix}
    0 & 1 & 0\\
    1 & 0 & 1\\
    0 & 1 & 0
  \end{smallmatrix}  \right).
  \end{cases}
$$
What is does can be interpreted as follows: if the local neighbourhood
is closer to  $\left( \begin{smallmatrix}
    1 & 0 & 1\\    
    0 & 1 & 0\\
    1 & 0 & 1
  \end{smallmatrix}  \right)$ 
  than to 
   $\left( \begin{smallmatrix}
    0 & 1 & 0\\
    1 & 0 & 1\\
    0 & 1 & 0
  \end{smallmatrix}  \right)$, it sets the central site to 1.
  If the local neighbourhood
is closer to 
   $\left( \begin{smallmatrix}
    0 & 1 & 0\\
    1 & 0 & 1\\
    0 & 1 & 0
  \end{smallmatrix}  \right)$
  than  to  $\left( \begin{smallmatrix}
    1 & 0 & 1\\    
    0 & 1 & 0\\
    1 & 0 & 1
  \end{smallmatrix}  \right)$,
  it sets the central site to 0. One could ask at this point what happens when the local neighbourhood is equally close to both  $\left( \begin{smallmatrix}
    1 & 0 & 1\\    
    0 & 1 & 0\\
    1 & 0 & 1
  \end{smallmatrix}  \right)$ and  $\left( \begin{smallmatrix}
    0 & 1 & 0\\
    1 & 0 & 1\\
    0 & 1 & 0
  \end{smallmatrix}  \right)$. The reader can easily verify that such situation cannot actually happen, thus we can simply write 
 $$
  f\begin{pmatrix}
     nw & n & ne\\
    w & c & e\\
    sw & s & se
  \end{pmatrix}  
  =\begin{cases}
  1 & 
  \text{if\,\,\,}
  H \left(\begin{smallmatrix}
       nw & n & ne\\
    w & c & e\\
    sw & s & se
  \end{smallmatrix},
  \begin{smallmatrix}
    1 & 0 & 1\\
    0 & 1 & 0\\
    1 & 0 & 1
  \end{smallmatrix}  \right) <
  H \left(\begin{smallmatrix}
       nw & n & ne\\
    w & c & e\\
    sw & s & se
  \end{smallmatrix},
  \begin{smallmatrix}
    0 & 1 & 0\\
    1 & 0 & 1\\
    0 & 1 & 0
  \end{smallmatrix}  \right),
   \\
  0 & \text{otherwise}.
  \end{cases} 
 $$ 
The definition of the above rule can be significantly simplified. 
Let $\overline{x}=1-x$. 
Note that
$$ H \left(\begin{smallmatrix}
       nw & n & ne\\
    w & c & e\\
    sw & s & se
  \end{smallmatrix},
  \begin{smallmatrix}
    1 & 0 & 1\\
    0 & 1 & 0\\
    1 & 0 & 1
  \end{smallmatrix}  \right)
=\overline{nw} +n+\overline{ne}
+w + \overline{c} + e
+ \overline{sw} +s+\overline{se},
  $$
and similarly
$$
 H \left(\begin{smallmatrix}
       nw & n & ne\\
    w & c & e\\
    sw & s & se
  \end{smallmatrix},
  \begin{smallmatrix}
    0 & 1 & 0\\
    1 & 0 & 1\\
    0 & 1 & 0
  \end{smallmatrix}  \right)
= {nw} +\overline{n}+{ne}
+\overline{w} + {c} + \overline{e}
+ {sw} +\overline{s}+{se}.
 $$   
Combining this together,
\begin{align*}
 H \left(\begin{smallmatrix}
       nw & n & ne\\
    w & c & e\\
    sw & s & se
  \end{smallmatrix},
  \begin{smallmatrix}
    1 & 0 & 1\\
    0 & 1 & 0\\
    1 & 0 & 1
  \end{smallmatrix}  \right)
  &-
   H \left(\begin{smallmatrix}
       nw & n & ne\\
    w & c & e\\
    sw & s & se
  \end{smallmatrix},
  \begin{smallmatrix}
    0 & 1 & 0\\
    1 & 0 & 1\\
    0 & 1 & 0
  \end{smallmatrix}  \right)=\\
&  \overline{nw} +n+\overline{ne}
+w + \overline{c} + e
+ \overline{sw} +s+\overline{se}\\
&-({nw} +\overline{n}+{ne}
+\overline{w} + {c} + \overline{e}
+ {sw} +\overline{s}+{se})\\
&=1-2nw -2ne -2c - 2sw-2se +2n+2w+2e+2s.
\end{align*}
This last expression will be negative when
$$
1 +2n+2w+2e+2s< 2nw +2ne +2c + 2sw+2se,$$
implying
$$
\frac{1}{2}+ n+w+e+s< nw +ne +c + sw+se.$$
All variables are 0 or 1, therefore the above equality is equivalent to
$$ n+w+e+s< nw +ne +c + sw+se,$$
leading to the simplified form of the function defining $f$,
$$
  f\begin{pmatrix}
     nw & n & ne\\
    w & c & e\\
    sw & s & se
  \end{pmatrix}  
=
\begin{cases}
1 & \text{if  $nw+ne+c+sw+se>n + w + e + s$,}\\
0 & \text{otherwise.}
\end{cases}
$$
Let $x \in \{0,1\}^{\mathbbm{Z}^2}$ be a configuration to be interpreted
as a grid of cells with integer coordinates $(i,j)$ for each cell, 
where each cell can take values $x_{i,j}\in \{0,1\}$.  
We define \emph{global function} $F$ associated with $f$ as
$$[F(x)]_{i,j}=
  f\begin{pmatrix}
     x_{i-1,j+1} &  x_{i,j+1} &  x_{i+1,j+1}\\
     x_{i-1,j} &  x_{i,j} &  x_{i+1,j}\\
     x_{i-1,j-1} &  x_{i,j-1} &  x_{i+1,j-1}
  \end{pmatrix}.
$$
We will consider a process of repeated consecutive applications of $F$ to some initial 
configuration $x$. In illustrations which follow we will represent 0's by white color and 1's by dark color
(blue or black).

\section{Single color damage}
The rule $F$ can repair some types of damages rather well. 
Let us suppose that we have infinite 2D checkerboard pattern configuration, to be denoted by $X^{\scalebox{0.7}{$\chboard$}}$,
 such that
$$
X^{\scalebox{0.7}{$\chboard$}}_{i,j}
=\begin{cases}
1 & \text{if $i+j$ is even}\\
0 & \text{otherwise.}
\end{cases}
$$
\begin{figure}
\begin{center}
\includegraphics[scale=0.23]{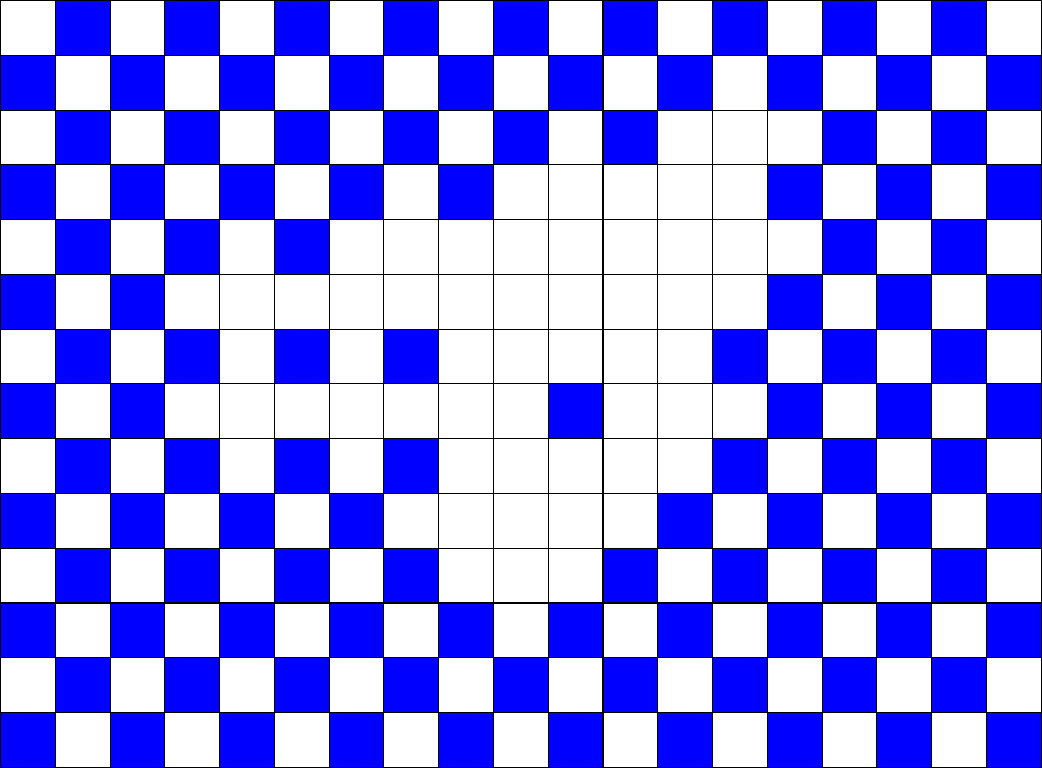}\,\raisebox{1.3cm}{\stackanchor{$F$}{$\longrightarrow$}}
\includegraphics[scale=0.23]{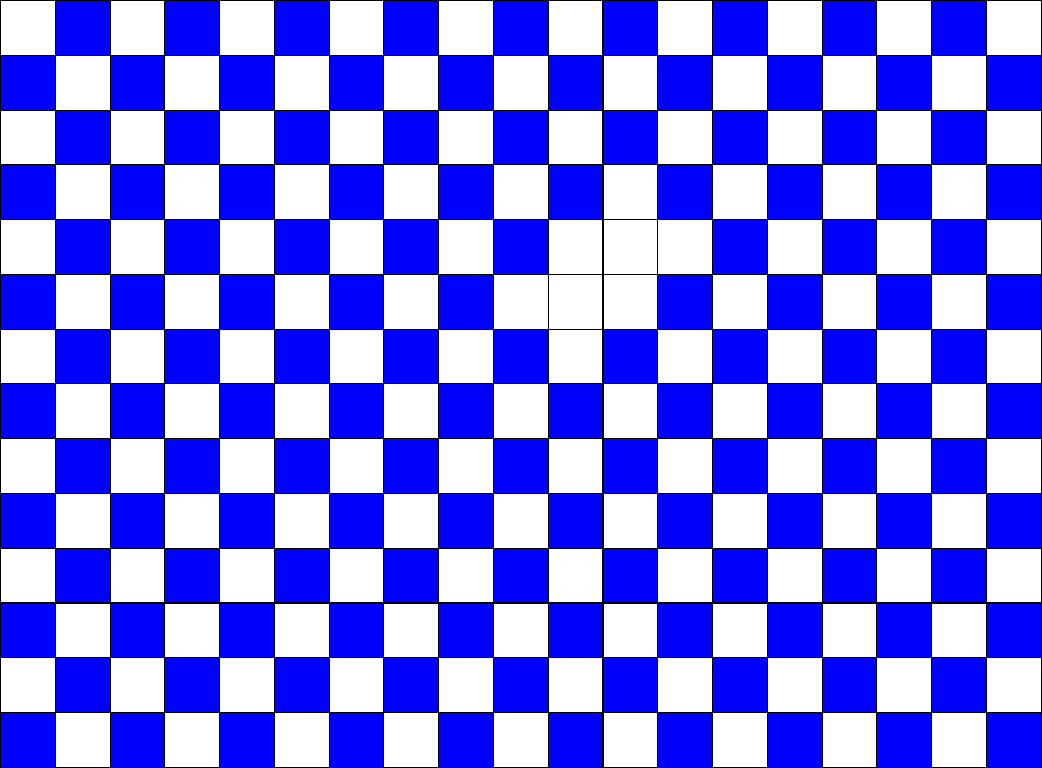}\,\raisebox{1.3cm}{\stackanchor{$F$}{$\longrightarrow$}}
\includegraphics[scale=0.23]{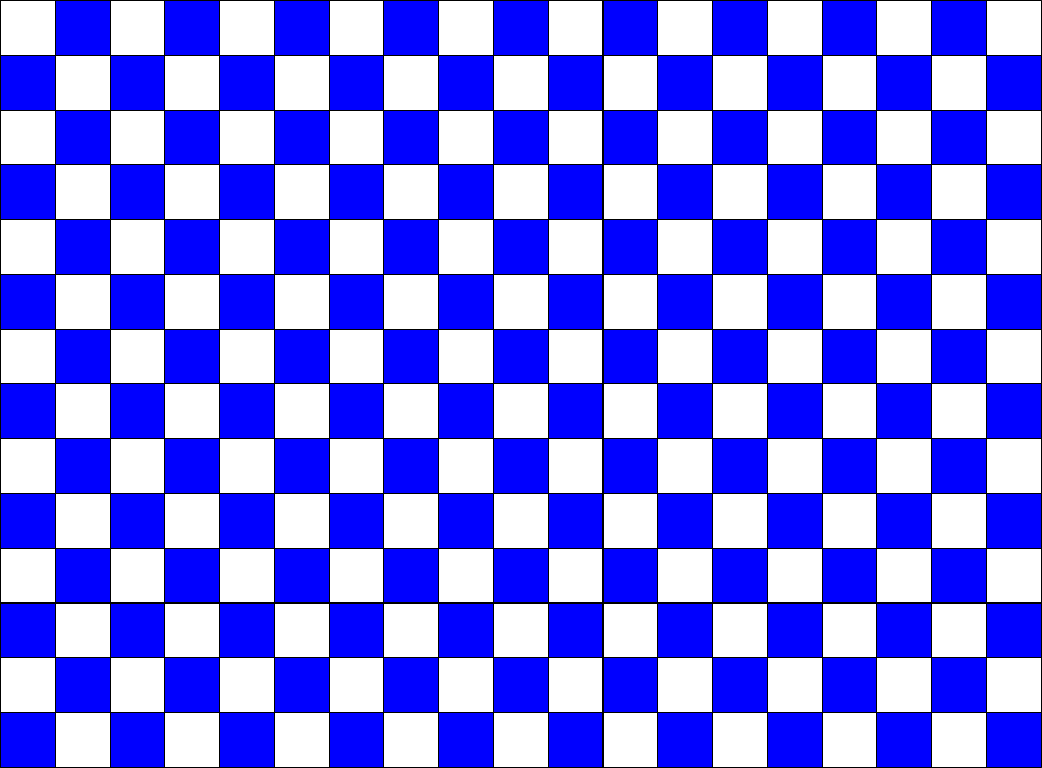}
\end{center}
\caption{Example of damage healing after removal of a finite number of 1's.}
\label{healexample}
\end{figure}
We will say that the  configuration is \emph{damaged} if it is obtained from $X^{\scalebox{0.7}{$\chboard$}}$ by modifying values of some
selected cells. If after a number of consecutive applications of the rule $F$
to such damaged patterns the  undamaged state $X^{\scalebox{0.7}{$\chboard$}}$
is restored, we will say
that the rule \emph{fixes the damage}.
\begin{proposition}
If the initial configuration $X^{\scalebox{0.7}{$\chboard$}}$ is damaged by replacing a finite number of  ones  by zeros, then  after a finite number of applications  of rule $F$ the damage will be fixed.
\end{proposition}
\emph{Proof.} Figure \ref{healexample} show example of the above type of damage.
In order to prove the proposition, let us first note that
all sites which are in state 1 in the initial damaged configuration will
stay in state 1 forever. This is because for each site in state 1, we have
$nw+ne+c+sw+se \geq 1$ and $n + w + e + s=0$, hence $nw+ne+c+sw+se>n + w + e + s$.

The number of damaged sites which were in state 1 in the original checkerboard pattern and after the damage assumed state 0 is finite. For all these sites
we have $n+w+e+s=0$. Among damaged sites, let us choose the one which has the smallest $i$ coordinate
(if there is more than one such site, choose the one with minimal $j$).
For this site we must have $nw+ne+c+sw+se>0$ and thus it will become 1 after
the  application of $F$. 
This means after each application of $F$ the number of damaged sites decreases at least by one, and eventually they disappear altogether restoring the original perfect
checkerboard pattern.
$\square$.

Analogous proposition holds for the removal of zeros, with a similar proof.
\begin{proposition}
If the initial configuration $X^{\scalebox{0.7}{$\chboard$}}$ is damaged by replacing a finite number of  zeros  by ones, then a after finite number of applications  of rule $F$ the damage will be healed.
\end{proposition}


\section{Two color damage}
Before considering a more general type of damage we will first
notice that the checkerboard voting rule is closely related to another
well known CA rule, namely the majority voting rule.
The local function of the  majority voting rule is defined as
$$
  g\begin{pmatrix}
     nw & n & ne\\
    w & c & e\\
    sw & s & se
  \end{pmatrix}  
=
\begin{cases}
1 & \text{if  $nw+w+ne+cw+c+ce+ sw+s+se>4$,}\\
0 & \text{otherwise.}
\end{cases}
$$
The corresponding global function will be denoted by $G$,
$$[G(x)]_{i,j}=
  g\begin{pmatrix}
     x_{i-1,j+1} &  x_{i,j+1} &  x_{i+1,j+1}\\
     x_{i-1,j} &  x_{i,j} &  x_{i+1,j}\\
     x_{i-1,j-1} &  x_{i,j-1} &  x_{i+1,j-1}
  \end{pmatrix}.
$$
Define transformation $T$  as 
$$
[T(x)]_{i,j}=x_{i,j}+ X^{\scalebox{0.7}{$\chboard$}}_{i,j} \mod 2
$$
Note that $T$ is a bijection (one-to-one). 
\begin{proposition}
The checkerboard voting rule $F$ and the majority voting rule $G$ are 
topologically conjugate, meaning that for every $x \in \{0,1\}^{\mathbbm{Z}^2}$,
$$G(T(x))=T(F(x)).$$
\end{proposition}
\begin{figure}
\begin{center}
$
\begin{CD}
 \raisebox{-0.5\height}{\includegraphics[scale=0.18]{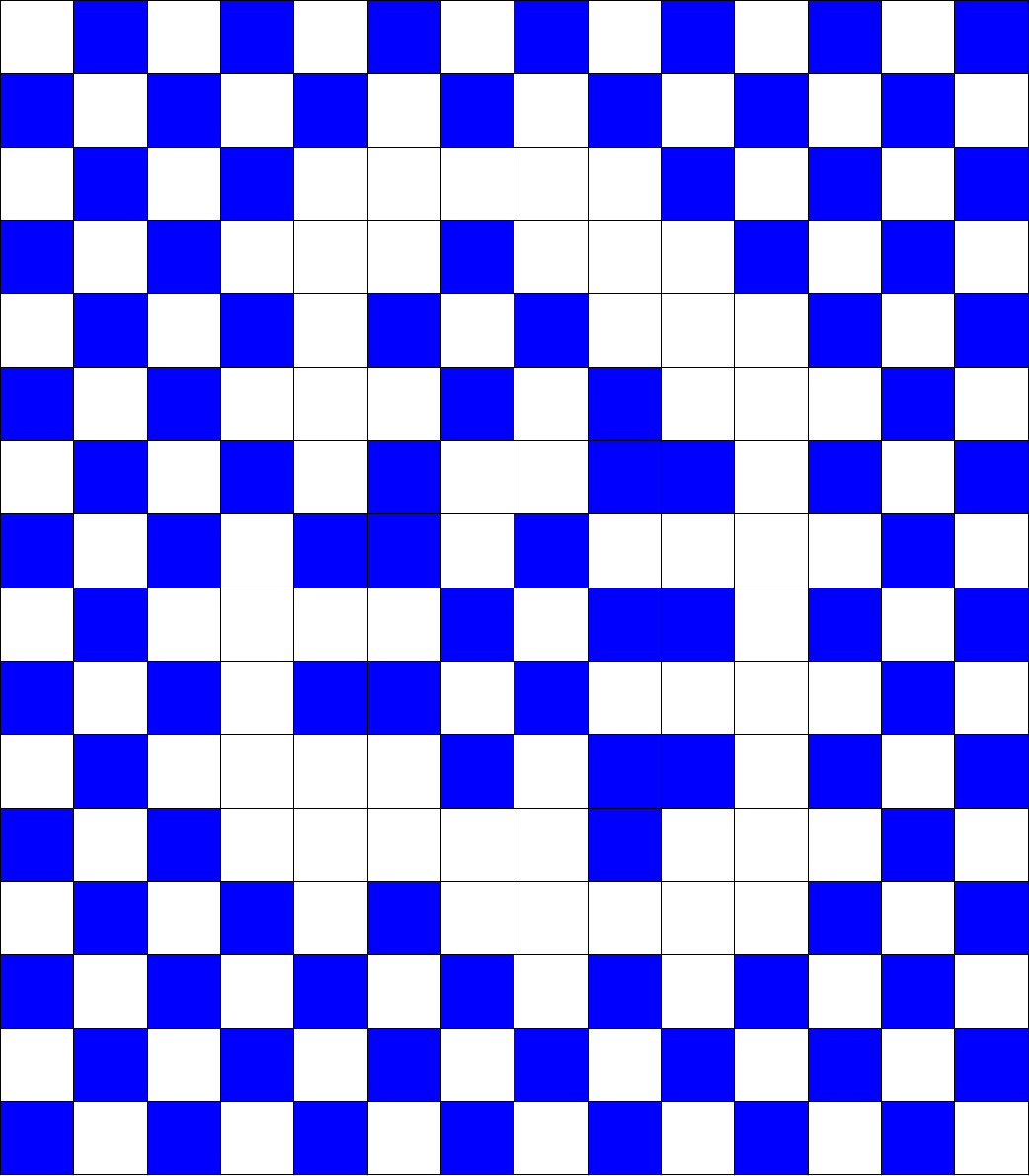}} @>T>> \raisebox{-0.5\height}{\includegraphics[scale=0.18]{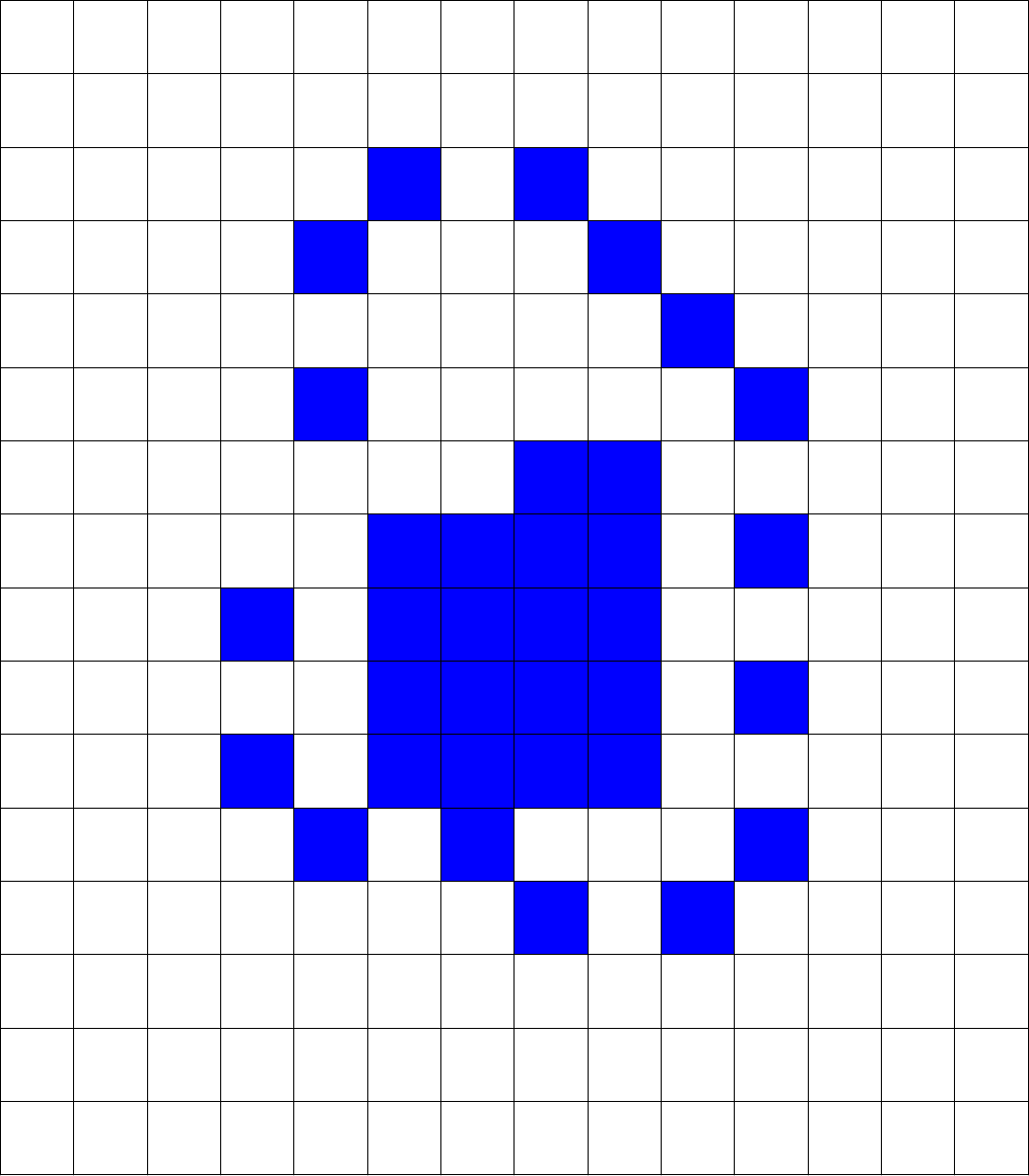}} \\
  @VVFV @VVGV \\
 \raisebox{-0.5\height}{\includegraphics[scale=0.18]{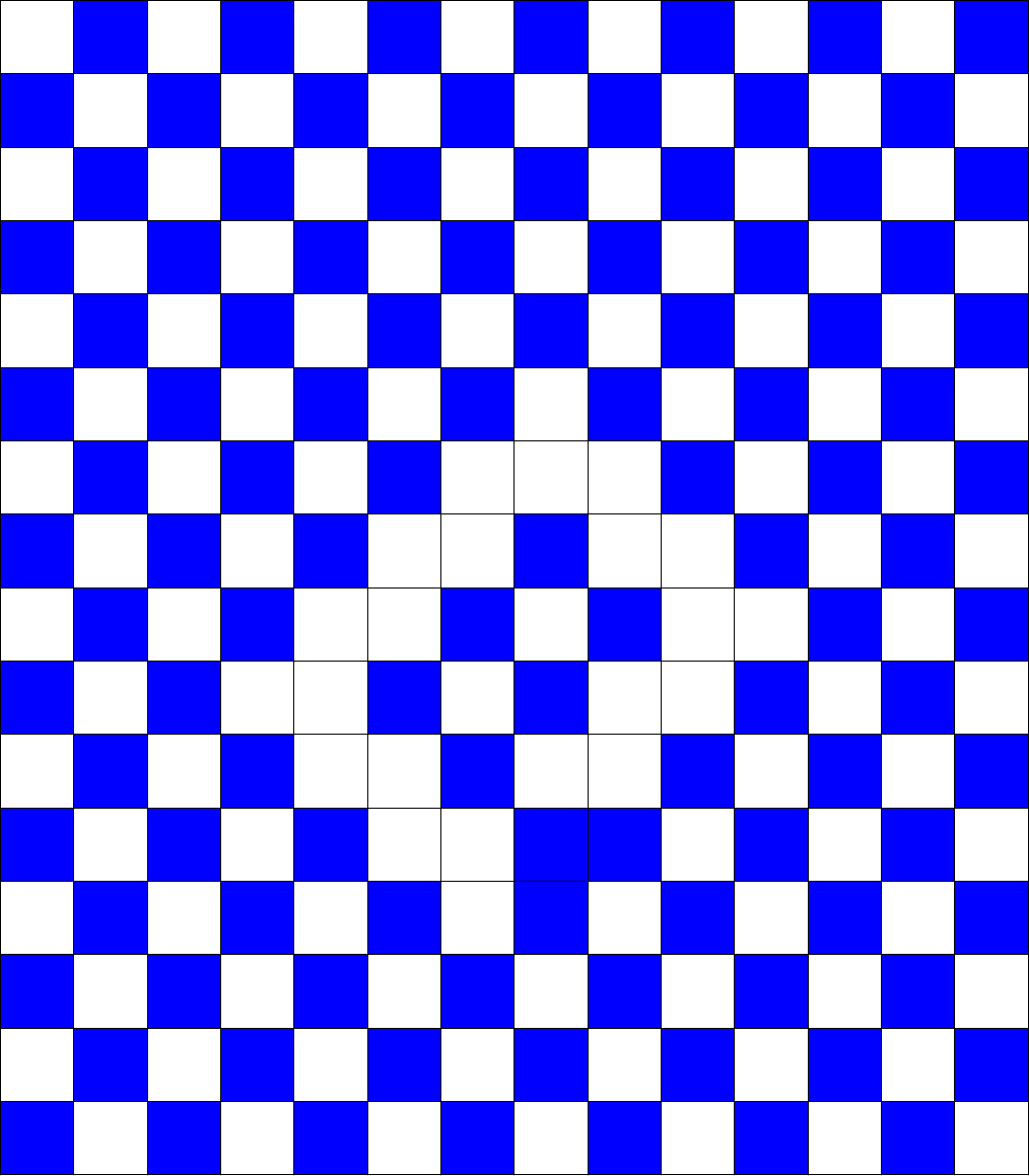}} @>T>> \raisebox{-0.5\height}{\includegraphics[scale=0.18]{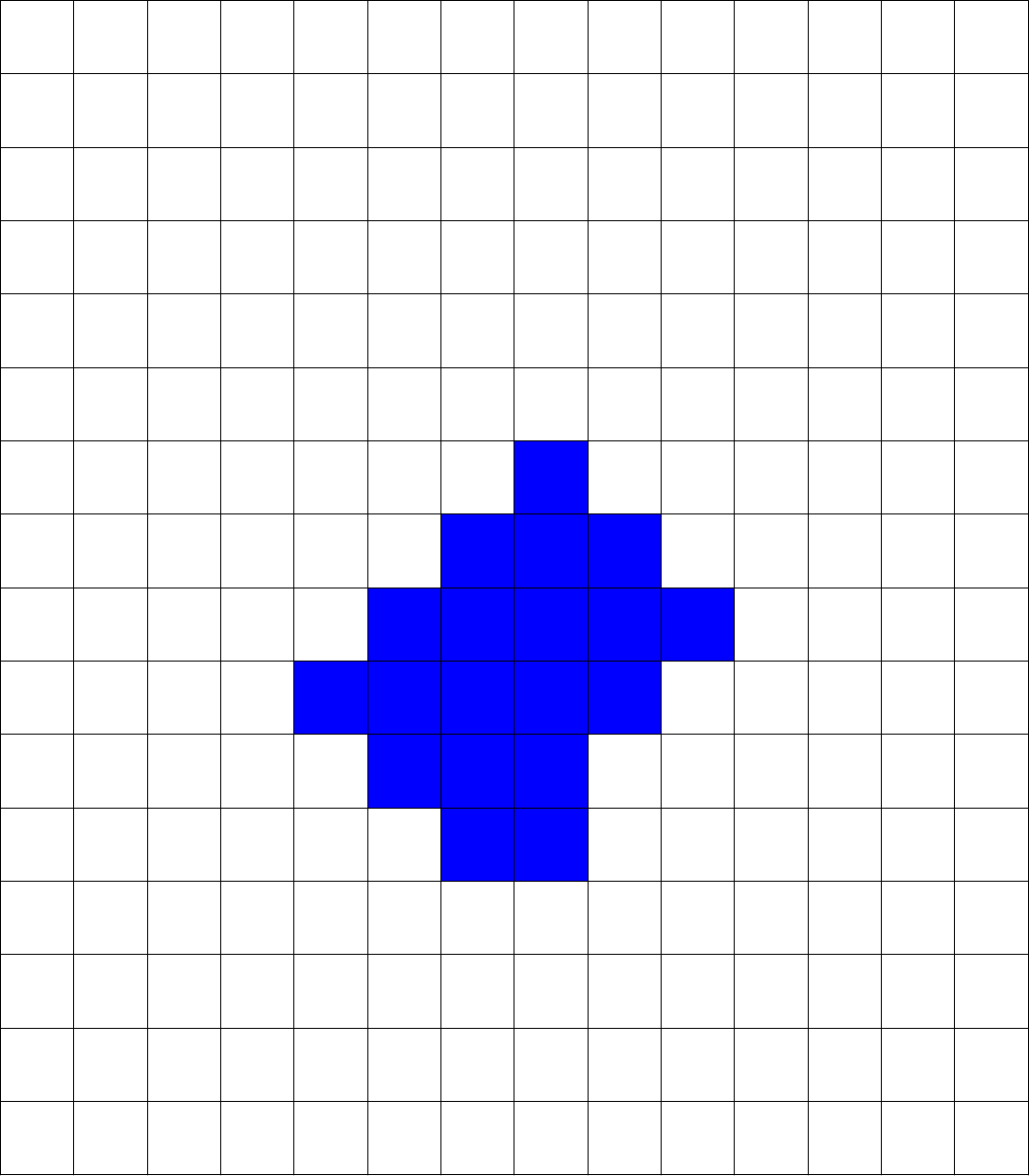}} \\
  @VVFV @VVGV \\
   \raisebox{-0.5\height}{\includegraphics[scale=0.18]{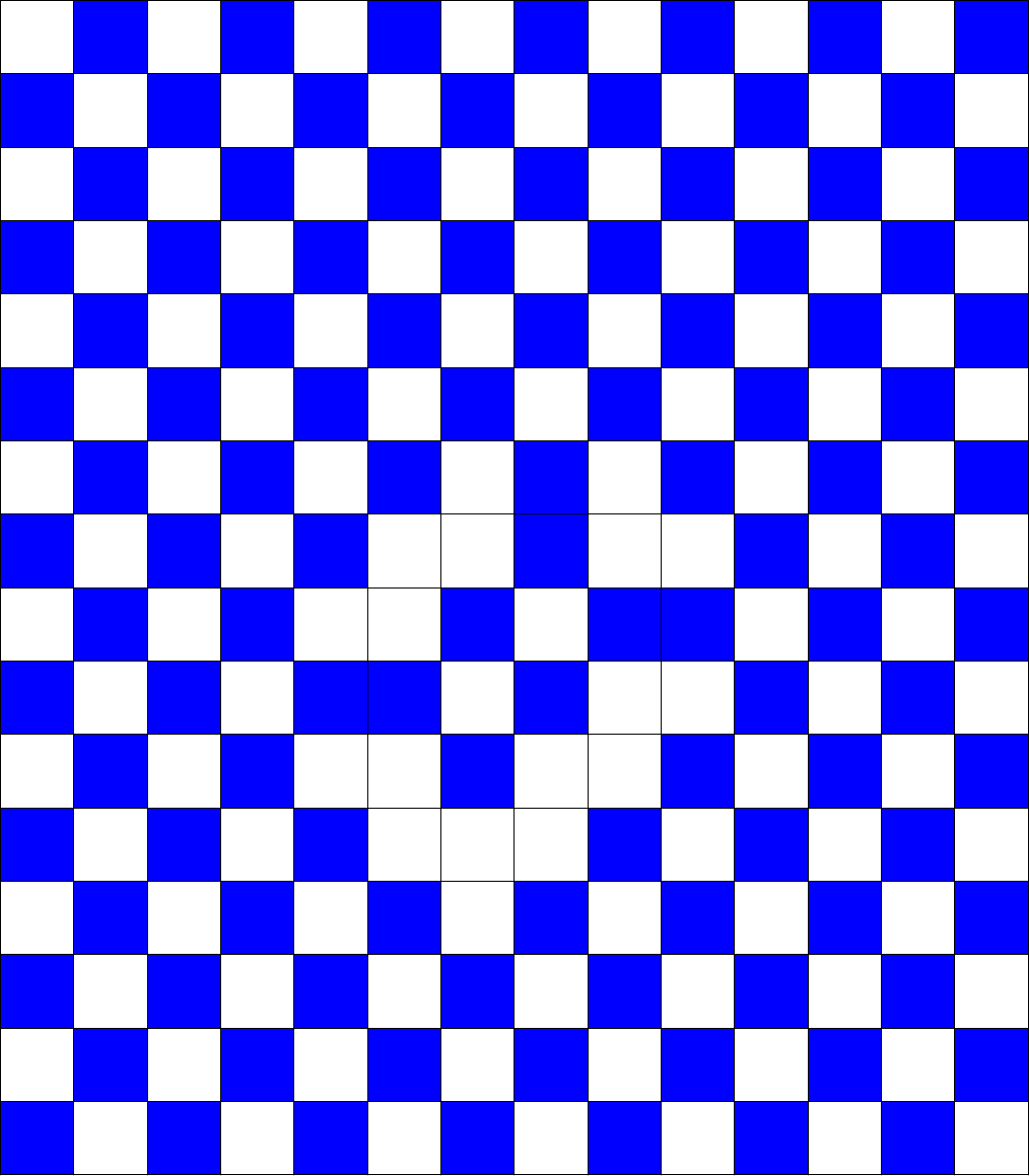}} @>T>> \raisebox{-0.5\height}{\includegraphics[scale=0.18]{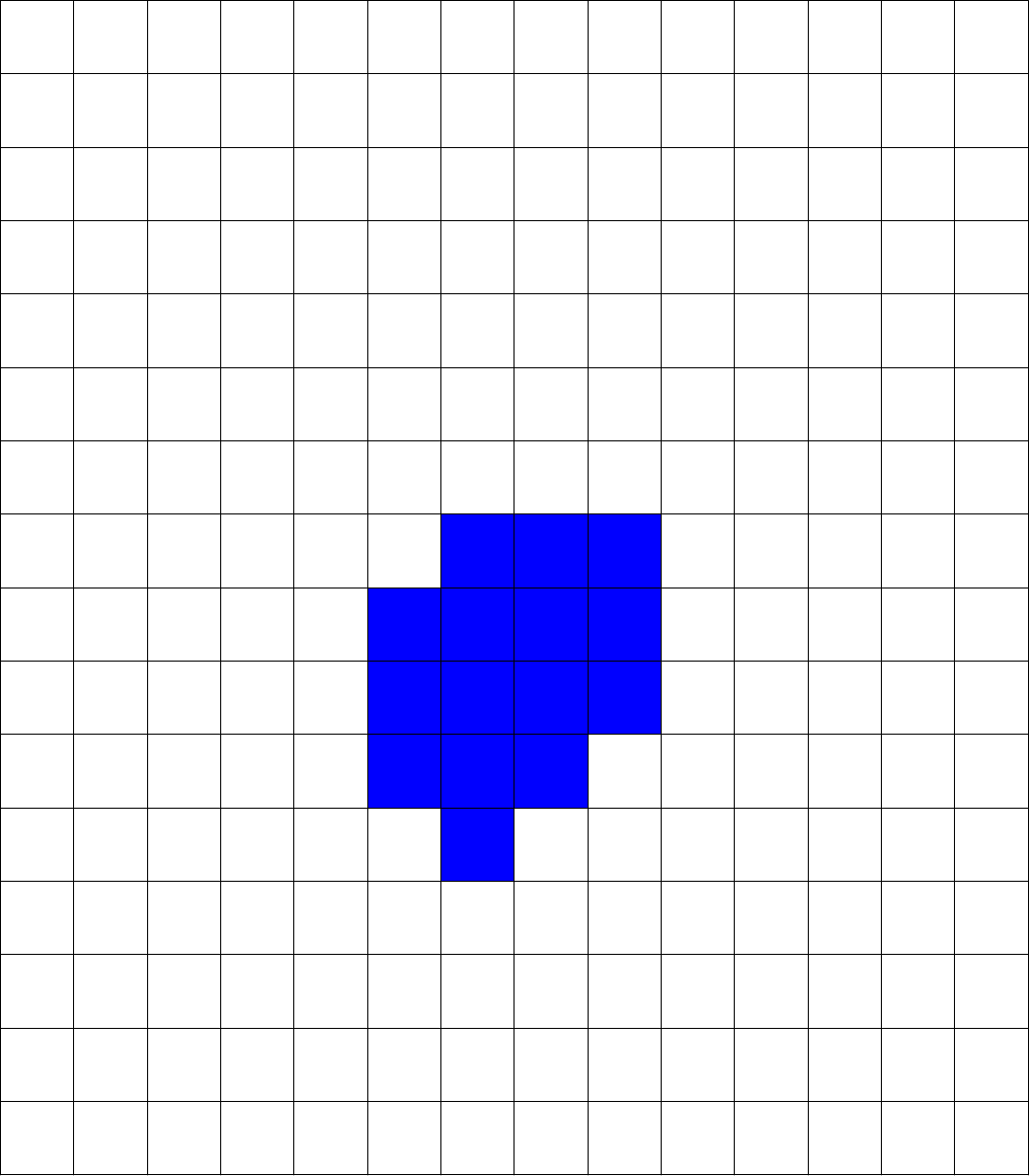}} \\
  @VVFV @VVGV \\
  \raisebox{-0.5\height}{\includegraphics[scale=0.18]{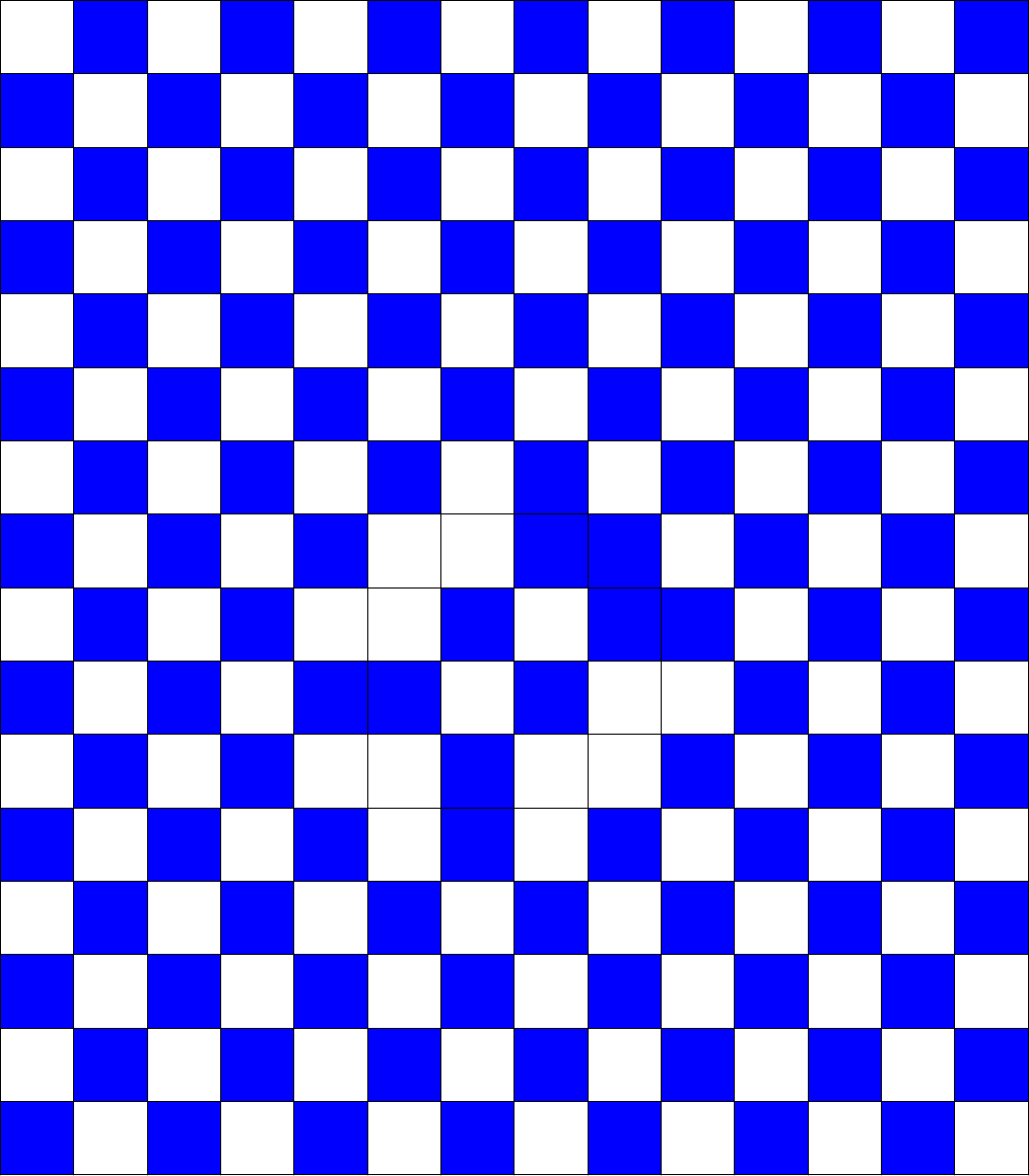}} @>T>> \raisebox{-0.5\height}{\includegraphics[scale=0.18]{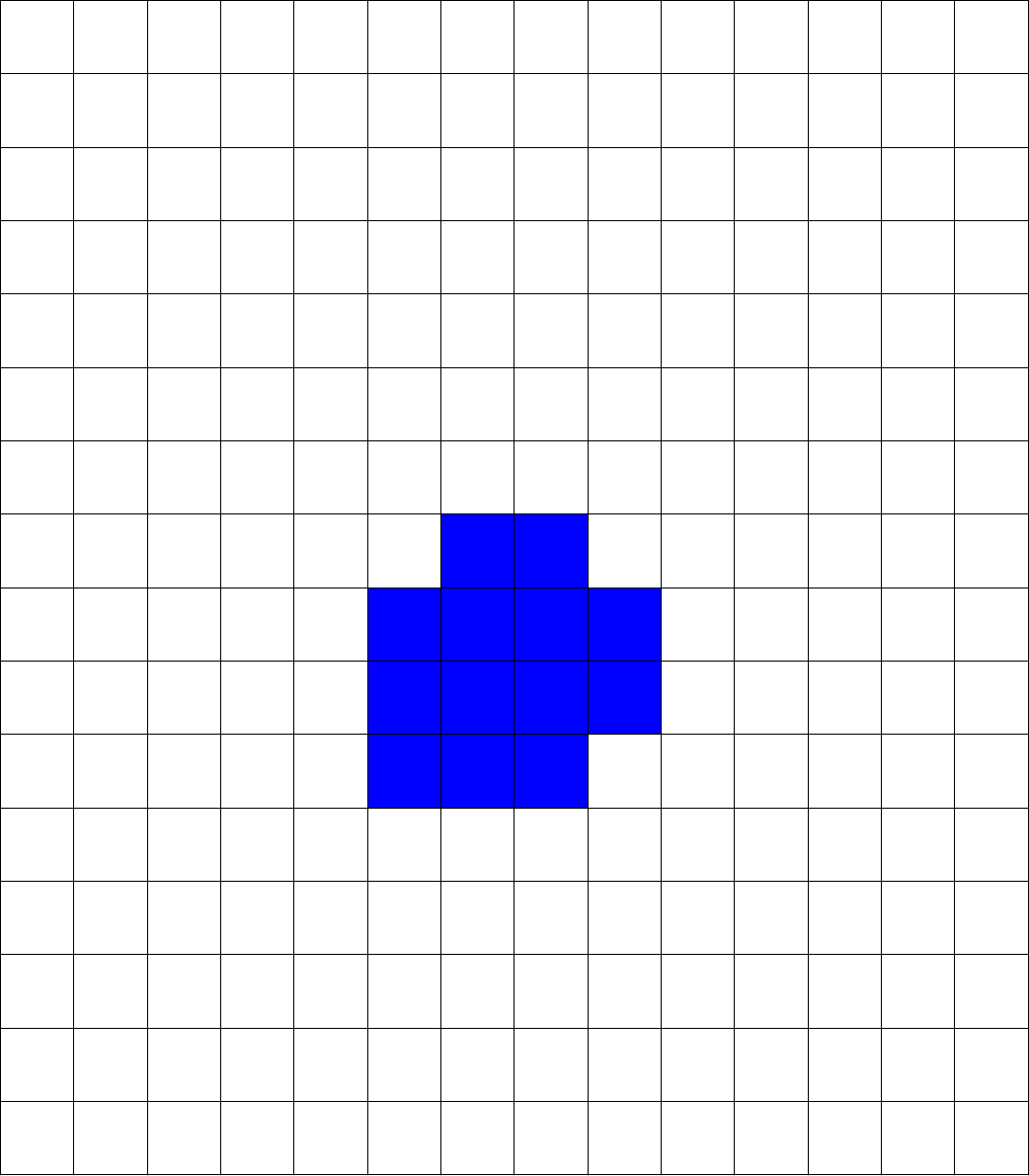}} \\
  @VVFV @VVGV \\efficiency-crop
  \raisebox{-0.5\height}{\includegraphics[scale=0.18]{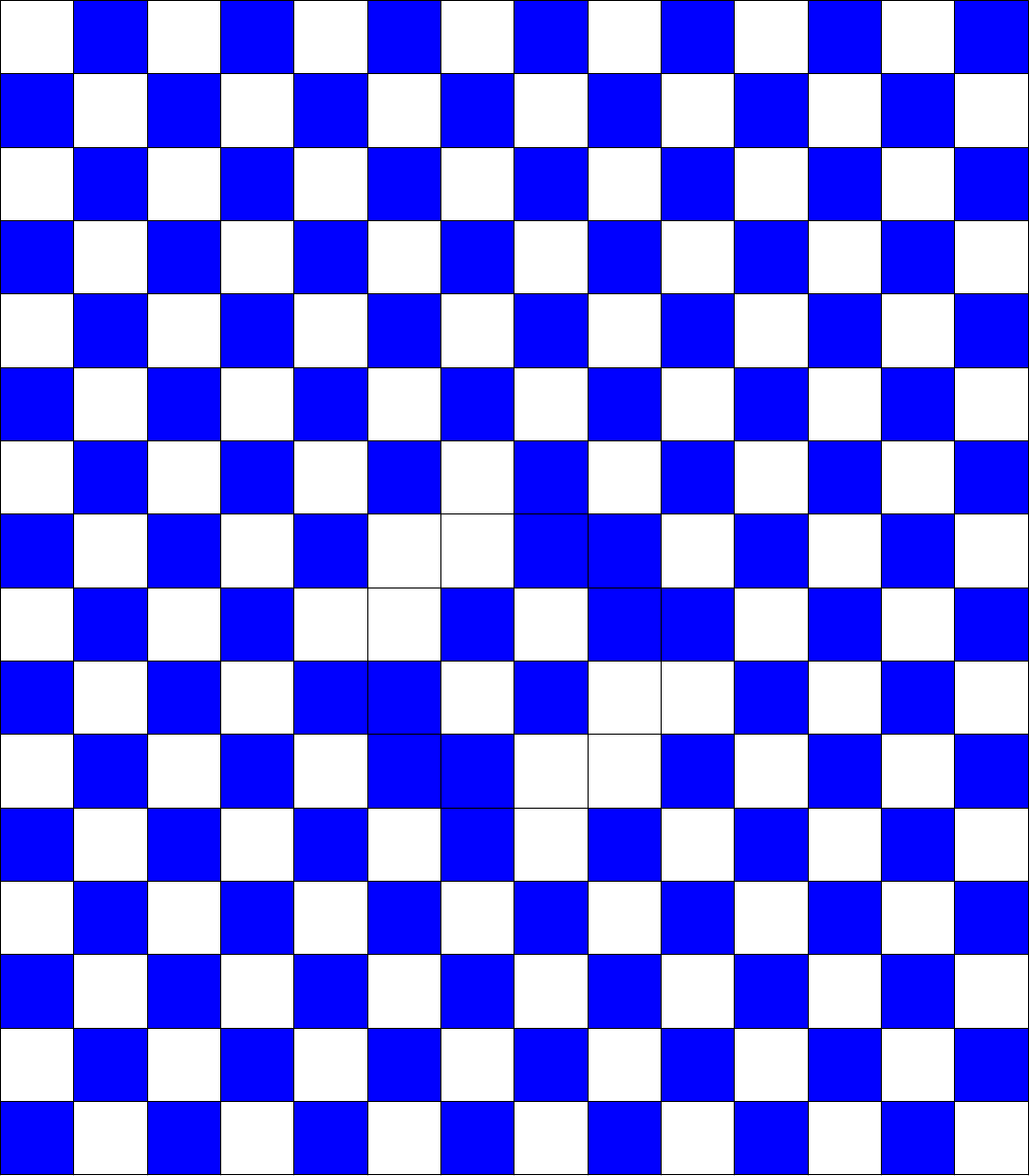}} @>T>> \raisebox{-0.5\height}{\includegraphics[scale=0.18]{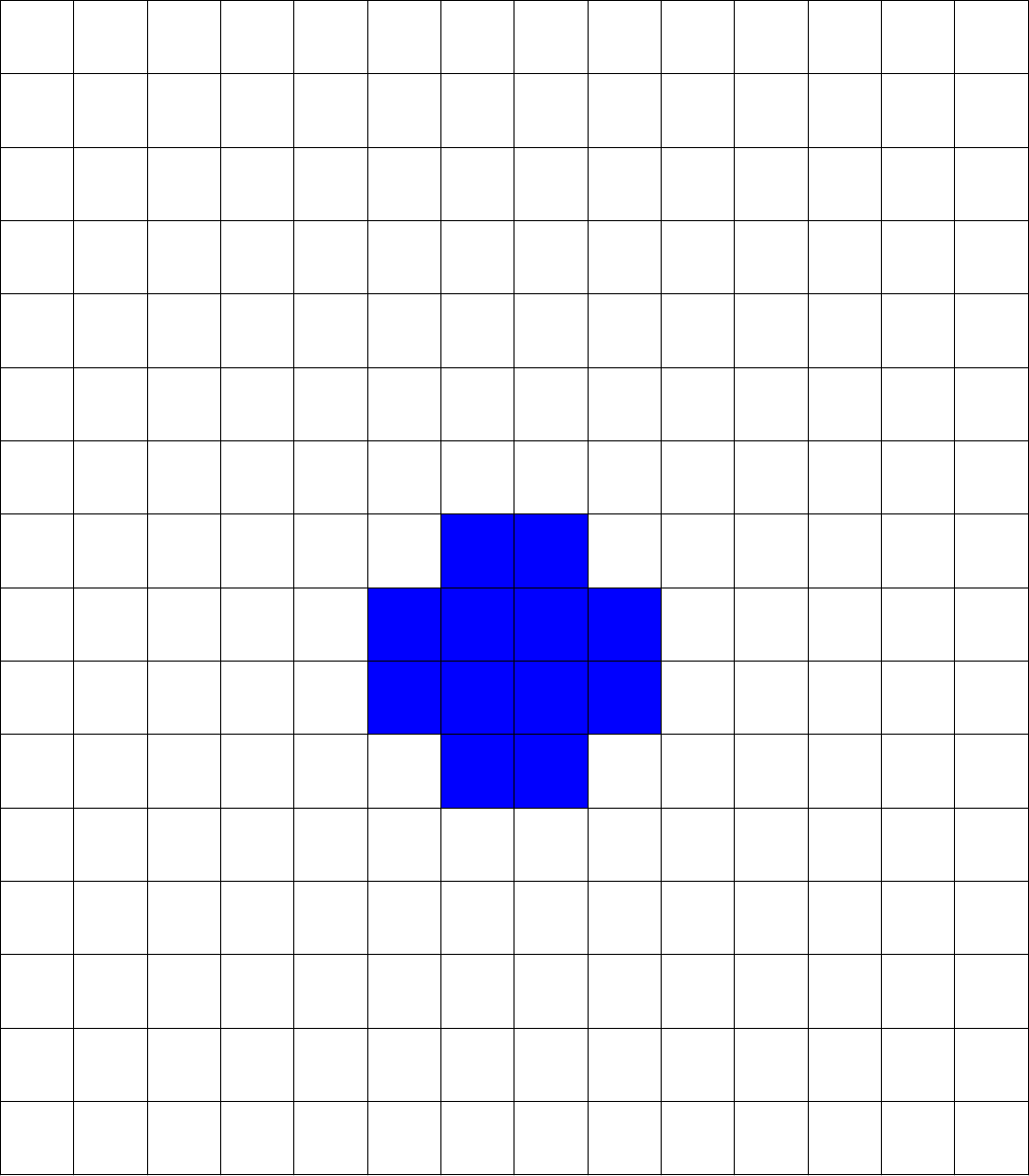}}
\end{CD}
$
\end{center}
\caption{Illustration of conjugacy between $F$ and $G$.}\label{conjugacyexample}
\end{figure}
Figure \ref{conjugacyexample} illustrates the conjugacy using a sample
damage which does not heal (the bottom configuration remains unchanged forever).
Note that if $x \in \{0,1\}^{\mathbbm{Z}^2}$ is obtained form 
$X^{\scalebox{0.7}{$\chboard$}}_{i,j}$ by damaging a finite number of sites, then
$T(x)$ will consist of all 0's except a finite number of 1's, and these 1's will be 
located in the same positions of $T(x)$ as damaged  sites in $x$.

\emph{Proof.} We will consider two cases. First suppose that  $i+j$ is even. Then
\begin{align}
[T(F(x)]_{i,j}&=[F(x)]_{i,j} + X^{\scalebox{0.7}{$\chboard$}}_{i,j} \mod 2
=[F(x)]_{i,j} +1 \mod 2\\ \nonumber
&=1-[F(x)]_{i,j}.
\end{align}
The above will be equal to 1 if and only if $[F(x)]_{i,j}=0$,
and this requires that
\begin{gather} \label{condition1}
{x}_{i+1,j-1}     
     +{x}_{i-1,j-1}
     +{x}_{i-1,j+1}
     +{x}_{i+1,j+1}
     +{x}_{i,j} 
     \leq
x_{i,j+1} +
     x_{i-1,j}  \\ +  x_{i+1,j}+
      +  x_{i,j-1}  \nonumber
\end{gather}

Now let $y=T(x)$. We have
$$[G(T(x)]_{i,j}= [G(y)]_{i,j}
= g\begin{pmatrix}
     y_{i-1,j+1} &  y_{i,j+1} &  y_{i+1,j+1}\\
     y_{i-1,j} &  y_{i,j} &  y_{i+1,j}\\
     y_{i-1,j-1} &  y_{i,j-1} &  y_{i+1,j-1}
  \end{pmatrix}.
$$
Because $i+j$ is even and $y=x+X^{\scalebox{0.7}{$\chboard$}} \mod 2$,
all values of $y$ on the diagonals of the above array will be Boolean conjugate of the corresponding $x$ values, while off-diagonal $y$ values will be the same as $x$,
$$[G(T(x)]_{i,j}=
 g\begin{pmatrix}
     \overline{x}_{i-1,j+1} &  x_{i,j+1} &  \overline{x}_{i+1,j+1}\\
     x_{i-1,j} &  \overline{x}_{i,j} &  x_{i+1,j}\\
     \overline{x}_{i-1,j-1} &  x_{i,j-1} &  \overline{x}_{i+1,j-1}
  \end{pmatrix}.
$$
Sum of the entries of the array is
\begin{gather*}\overline{x}_{i-1,j+1} +  x_{i,j+1} +  \overline{x}_{i+1,j+1}+
     x_{i-1,j} +  \overline{x}_{i,j} +  x_{i+1,j}+
     \overline{x}_{i-1,j-1} +  x_{i,j-1} +  \overline{x}_{i+1,j-1}\\
 =  5 +  x_{i,j+1} +
     x_{i-1,j}  +  x_{i+1,j}+
      +  x_{i,j-1} -{x}_{i+1,j-1}  \\
     -{x}_{i-1,j-1}
     -{x}_{i-1,j+1}
     -{x}_{i+1,j+1}
     -{x}_{i,j}
     \end{gather*}
$[G(T(x)]_{i,j}$ will be equal to 1 when the above is greater than 4, i.e.,  when 
\begin{gather*}
1 +  x_{i,j+1} +
     x_{i-1,j}  +  x_{i+1,j}+
      +  x_{i,j-1}  \nonumber \\
     >
     {x}_{i+1,j-1}     
     +{x}_{i-1,j-1}
     +{x}_{i-1,j+1}
     +{x}_{i+1,j+1}
     +{x}_{i,j}.
\end{gather*}
Because we are dealing with positive integers, this is equivalent to 
\begin{gather} \label{condition2}
  x_{i,j+1} +
     x_{i-1,j}  +  x_{i+1,j}+
      +  x_{i,j-1}  
     \geq
     {x}_{i+1,j-1}     
     +{x}_{i-1,j-1}
     +{x}_{i-1,j+1} \\
     +{x}_{i+1,j+1}
     +{x}_{i,j}. \nonumber
\end{gather}
Note that conditions (\ref{condition1}) and (\ref{condition2}) are identical, proving that
$[G(T(x))]_{i,j}=[T(F(x))]_{i,j}$, as required.
The case of odd $i+j$ is very similar, thus we will not repeat it here.
$\square$

In \cite{Gartner21}, B. Gärtner and  A. N. Zehmakan studied various properties
of the majority voting rule. One of their results is stated below (Lemma 3.14
in their paper,
see  \cite{Gartner21} for proof).
\begin{proposition}
If the initial configuration contains at most 11 cells in state 1, after a finite number of iterations of the majority voting rule $G$ the process will reach a configuration with all cells in state 0.
\end{proposition}
The conjugacy between $F$ and $G$ allows us to immediately rephrase the above result for $F$.
\begin{proposition}\label{proposition12sites}
If $x \in \{0,1\}^{\mathbbm{Z}^2}$ is obtained form 
$X^{\scalebox{0.7}{$\chboard$}}_{i,j}$ by damaging at most 11 sites,
then after a finite number  of iterations of the checkerboard voting rule $F$ the process will reach the perfect checkerboard pattern $X^{\scalebox{0.7}{$\chboard$}}_{i,j}$.
\end{proposition}
Note that the damage can be of arbitrary type, meaning that some 0's may be replaced by 1's and some 1's by 0's, as long as the total number of damaged cells is not more than 11.
If the total number of damaged sites is 12 or more, the pattern may or may not heal,
but there seem to be no obvious characteristics of the damage which could determine
\emph{a priori} whether the patterns heals or not. For this reason, 
we will study the problem form a numerical perspective and propose a measure of
efficiency of damage healing.

\section*{Healing efficiency} 
Let us suppose that we start with an infinite 2D checkerboard pattern configuration $X^{\scalebox{0.7}{$\chboard$}}$.
Now we select $n\times n$ square of sites with indices $0 \leq i \leq n-1$ and 
$0 \leq j \leq n-1$ and flip some selected bits in this square (by ``flipping'' 
site $(i,j)$ we mean replacing $x_{ij}$ by $\overline{x}_{i,j}$).
This will be called \emph{square damage of order $n$}. If there exists a finite $m$ such that after $m$ iterations of the CA rule $F$ such damage is healed and we obtain the original perfect checkerboard pattern $X^{\scalebox{0.7}{$\chboard$}}$, we call the damage \emph{fixable}. 

It is clear that the total number of possibilities of square damage of order $n$
is $2^{n^2}$, because we have $n^2$ possible bits to flip. We can define \emph{healing efficiency for square damage of order $n$} as
$$
\chi(n)=\frac{\text{ \# of fixable damages of order $n$}}{2^{n^2}}.$$
\begin{proposition}
All square damages of order 3 or less are fixable, meaning that
$$\chi(n)=1$$ for all $n\leq 3$.
\end{proposition}

This results is a direct consequence of Proposition~\ref{proposition12sites} which says that any damage of up to 11 sites is fixable. Since in squares $1 \times 1$,
$2 \times 2$ and $3 \times 3$ there are less than 11 sites, all square damages  of order 1, 2 and 3 are fixable.
\begin{figure}
\begin{center}
\includegraphics[scale=0.82]{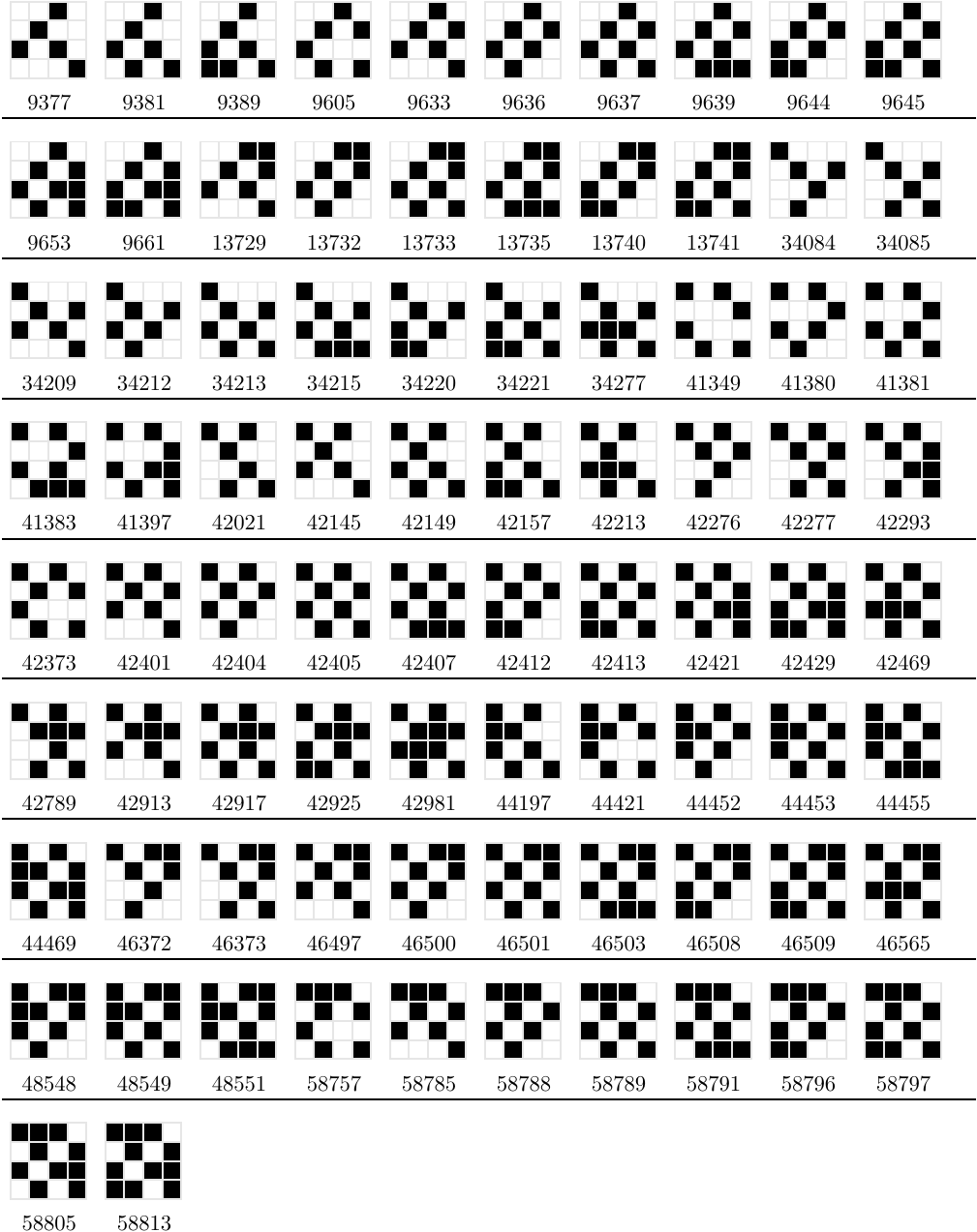}
\end{center}
\caption{List of 82 square damages $4 \times 4$ which are not fixable.
Numbers represent decimal coding of 16-bit  binary strings
corresponding to cell values.} \label{82noheal}
\end{figure}

For $n=4$, we checked all possible
$2^{16}=65536$ damages using a computer program. Among them, 82 do not heal and the system reaches a fixed point which is different from the perfect checkerboard configuration.
The remaining 65454 damages are fixable, yielding
$$
\chi (4)=\frac{65454}{65536} \approx 0.9987,
$$
which is still remarkably close to 1. It is, nevertheless, instructive to
take a closer look at the 82 damages which do not heal.
They are shown in Figure~\ref{82noheal} with numbers below which represent decimal coding of 16-bit  binary strings corresponding to cell values. All of these
damages after one application of $F$ yield configuration 13740, which is a ``fixed
point'' (does not change under the action of $F$).
 Furthermore, all of them have at least 12 cells damaged, as one would expect from Proposition \ref{proposition12sites}.
The smallest number (12) of damaged cells  can be found in the ``fixed
point''  configuration 13740, where all cells are damaged except of four corners.
The bottom left of Figure~\ref{conjugacyexample} shows the configuration 13740
embedded into the checkerboard pattern, and bottom right of the same figure shows
the corresponding $T(x)$. One can see that $T(x)$ consist of 12
cells arranged in a $4 \times 4$ square with four corners removed. 
All remaining configurations shown in 
Figure~\ref{82noheal}  have 13, 14, 15 or 16 cells damaged, thus their images
under $T$ would have also, correspondingly,  13, 14, 15 or 16 cells in state 1 surrounded by all zeros.

When $n=5$, the number of damages which do not heal is 596012. In contrast to
$n=4$ case, they do not converge to the same single fixed point, but rather to 15 
different fixed points. Efficiency in this case is 
$$\chi(5)=\frac{32958420}{2^{25}} \approx 0.9822.$$

\begin{figure}
\begin{center}
\includegraphics[scale=1.0]{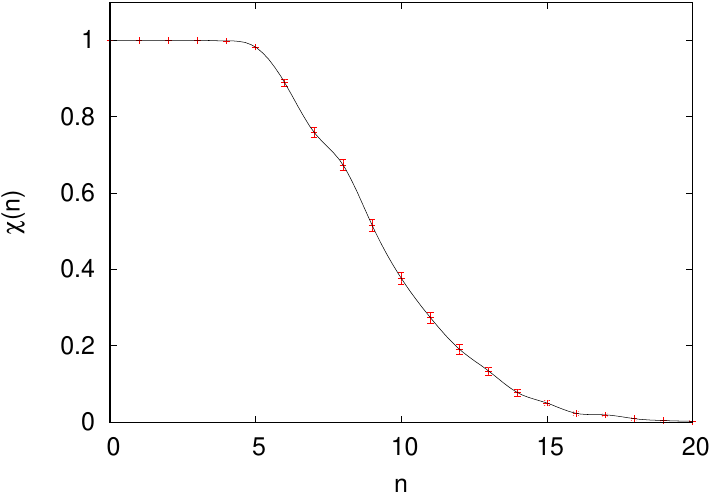}
\end{center}
\caption{Estimates of $\chi(n)$ by Monte Carlo method.
Error bars represent standard deviation. Values of $\chi(n)$ for $n=1,2,3,4,5$ are exact.}\label{figurechi}
\end{figure}

As $n$ increases even further, it becomes impossible to simply check all damages 
whether they fix or not, due to the fact that their number grows 
as $2^{n^2}$. Nevertheless, one can estimate $\chi (n)$ by Monte Carlo
method. If  $N$ damages are generated randomly from uniform probability distribution
and $N_h$ of them heal, then, by the law of large numbers
$$ \chi(n) =\lim_{N \to {\infty}} \frac{N_h}{N},
$$
thus 
$$ \chi(n) \approx \frac{N_h}{N}
$$
if $N$ is large. A simple way to sample damages from uniform probability distribution
is to set bits inside $n \times n$ square to 0 or 1 with probability 0.5,
independently of each other.
Results of such experiments are shown in Figure~\ref{figurechi}. 
We can clearly see that the efficiency quickly diminishes  with increasing $n$.

\begin{proposition}
The healing  efficiency of the checkerboard voting rule   tends to zero as $n\to \infty$,
$$\lim_{n \to \infty}  \chi (n) =0.$$ 
\end{proposition}

\emph{Proof.} Consider $4 \times 4$ damage  configuration 13740 shown in
 in Figure~\ref{82noheal}. Define set $S$ of all cells of this configuration
 except corners, 
 $$S=
\begin{matrix}
      & 0 &  1 &   \\
    0 & 1 & 0 & 1 \\
    1 & 0 & 1 & 0\\
      & 1 & 0 &    
  \end{matrix} 
 $$
 It is easy to verify that all cells belonging to $S$ 
 will remain in the same state after application of $F$, regardless of what
 lies outside of $S$. We can call it a \emph{permanent set}.
 If $S$ appears somewhere in a given $n \times n$ damage, the damage will
 not heal. 
 Let the probability that the permanent set $S$ appears in a randomly selected
$4\times 4$ square be called $p$. The value of $p$ can be easily computed -- there are $2^4$ 
possible configurations of the four missing corners, thus we have
$p= 2^4/2^{16}=2^{-12}.$
 
For a given large $n$, divide $n \times n$ square into $4 \times 4$ sub-squares,
leaving a small margin on the left and in the bottom in case when $n$ is not divisible by 4. We will have $\lfloor n/4 \rfloor^2$ such sub-squares. In a randomly selected
$n \times n$ configuration, the probability that $S$ does not appear in any of the
aforementioned $4 \times 4$ sub-squares  is $(1-p)^{\lfloor n/4 \rfloor^2}$.
The probability that it does appear in at least one of the subsquares, therefore,
is $1-(1-p)^{\lfloor n/4 \rfloor^2}$. The probability that the $n \times n$ square damage does not heal is larger than this number, because, among other reasons, $S$ can appear
also in a $4\times 4$ region overlapping two subsquares. Probability of
non-healing is $1 - \chi(n)$, hence
$$1 - \chi(n) > 1-(1-p)^{\lfloor n/4 \rfloor^2}.$$
From this we conclude that
$$0<\chi(n) < (1-p)^{\lfloor n/4 \rfloor^2},$$
and, by the squeeze theorem, $\chi(n) \to 0$ as $n \to \infty$. $\square$

\section*{Conclusion and further questions}
We introduced a simple cellular automaton model for a self-repairing system and explored its characteristics. In this model, the underlying structure is represented by a two-dimensional checkerboard arrangement, which can sustain damage by altering the values at a limited number of sites. The cellular automaton rule used in the model is a checkerboard voting rule which is topologically equivalent to the majority voting rule. When damage occurs with a single color, the rule consistently restores the affected areas. For localized damage within a $3 \times 3$ square, the rule also reliably repairs it. However, when damage occurs within a larger $n \times n$ square, the rule's effectiveness in restoring the area drops below 100\%, yet it remains over 98\% for $n \leq 5$ and above 75\% for $n \leq 7$. We demonstrated that as $n$ approaches infinity, the efficiency of damage repair ultimately trends towards zero.

There is a number of interesting questions which remain. 
The most important one is the following: are there any better rules, with better efficiency? We suspect that among CA rules in $3 \times 3$ Moore neighbourhood there are no better ones, but we have no proof for this yet. 
On can also ask a more general question: is there a rule (of any neighbourhood size) for which $\chi=1$ for all $n$? Again, we suspect that no such rule exists, but the work exploring this issue is still ongoing.


\begin{thebibliography}{1}
\providecommand{\url}[1]{\normalfont{#1}}
\providecommand{\urlprefix}{Available from: }

\bibitem{KELLER2018431}
Keller~MW, Crall~MD. Self-healing composite materials. In: Beaumont~PW,
  Zweben~CH, editors. Comprehensive composite materials II. Oxford: Elsevier;
  2018. p. 431--453.

\bibitem{Kutten99a}
Kutten~S, Peleg~D. Fault-local distributed mending. Journal of Algorithms.
  1999;\hspace{0pt}30(1):144--165.

\bibitem{Kutten99b}
Kutten~S, Peleg~D. Tight fault locality. SIAM Journal on Computing.
  2000;\hspace{0pt}30(1):247--268.

\bibitem{Peleg2002}
Peleg~D. Local majorities, coalitions and monopolies in graphs: a review.
  Theoretical Computer Science. 2002;\hspace{0pt}282(2):231--257.

\bibitem{Gartner21}
Gärtner~B, Zehmakan~AN. Majority rule cellular automata. Theoretical Computer
  Science. 2021;\hspace{0pt}889:41--59.

\end{thebibliography}
\end{document}